# Enabling and Optimizing Pilot Jobs using Xen based Virtual Machines for the HPC Grid Applications

Omer Khalid
CERN
Geneva, Switzerland
Omer.Khalid@cern.ch

Richard Anthony
University of Greenwich
London, United Kingdom
R.J.Anthony@gre.ac.uk

Paul Nilsson
University of Texas
Texas, USA
Paul.Nilsson@cern.ch

Kate Keahey
Argonne National Laboratory
Chicago, USA
Keahey@mcs.anl.gov

Markus Schulz
CERN
Geneva, Switzerland
Markus.Schulz@cern.ch

Kevin Parrott
University of Greenwich
London, United Kingdom
A.K.Parrott@gre.ac.uk

Miltos Petridis
University of Greenwich
London, United Kingdom
M.Petridis@gre.ac.uk

## ABSTRACT
The primary motivation for uptake of virtualization have been resource isolation, capacity management and resource customization: isolation and capacity management allow providers to isolate users from the site and control their resources usage while customization allows end-users to easily project the required environment onto a variety of sites. Various approaches have been taken to integrate virtualization with Grid technologies. In this paper, we propose an approach that combines virtualization on the existing software infrastructure such as Pilot Jobs with minimum change on the part of resource providers. We also present a standard API to enable a wider set of applications including Batch systems to deploy virtual machines on-demand as isolated job sandboxes. To illustrate the usefulness of this approach, we also evaluate the impact of Xen virtualization on memory and compute intensive tasks, and present our results that how memory and scheduling parameters could be tweaked to optimize job performance.

## Categories and Subject Descriptors
D.2.7 [**Software Engineering**]: Distribution, Maintenance, and Enhancement – *portability*; Metrics – *performance measures*;

## 1. INTRODUCTION
Deployment of virtualization on the Grid [2] has been an active arena of research since the virtualization technology became an established desktop tool [23, 24]. This is particularly challenging for large research organizations like European Organization for the Nuclear Research (CERN) where compute nodes have to be configured for large set of software environments: if a platform can be configured by simply deploying a virtual machine (VM) rather than arduously reconciling the needs of several different communities in one static configuration, the site has a better chance to serve multiple communities. The work we present here is to address part of that problem, and we propose a solution to tackle it.

Given this potential, we decided to investigate how this technology could benefit ATLAS [21] (one of CERN's high energy physics experiments) grid infrastructure and improve its efficiency. ATLAS computing framework[1] is a high level software layer over the grid middleware that manages job submission, job monitoring, job resubmission in case of failures, queue management with different scientific grids such as OSG[2], LCG[3], NorduGrid[4] while providing a coherent interface to its users. Specifically, we focus on the PanDA system [20], which is based on the pilot job submission framework [21].

---

[1] More information is available from this website: http://www.gridpp.ac.uk/eb/ComputingModels/

[2] Open Science Grid: http://www.opensciencegrid.org

[3] Large Hadron Collider Computing Grid: http://www.cern.ch/lcg

[4] NorduGrid: http://www.nordugrid.org

The ATLAS computing framework uses two different submissions mechanisms as described next.

### 1.1.1 Direct submission
In this model, ATLAS scientific community users submit their jobs directly to a specific grid. Usually, a user submits to the Workload Management Systems (WMS), which then provides resource matchmaking and submission to a particular site. This could lead to substantial load on the WMS under peak usage and introduces scheduling latencies, which reduces its efficacy and overall throughput as failing jobs end up at the back of job queue and are re-submitted.

### 1.1.2 Leasing Job Slots
In recent years, ATLAS computing framework has developed job submission frameworks that interfaces with different scientific grids and takes cares of job submission and monitoring. The present PanDA system submits pilot jobs, acquire resources slots on the ATLAS grid sites and then downloads the jobs for execution from high-level ATLAS computing framework queues. Pilot jobs are a kind of resource leasing mechanism, and are responsible for delegating the job scheduling responsibility from the workload management systems to PanDA Server yielding improved resource provisioning and allocation for ATLAS jobs.

In section 3, propose an approach to introduce virtualization at grid computer nodes. In section 4, we evaluate a case study and measure performance metrics of ATLAS jobs by executing them in VM. Later on, we discuss the on-going research projects, which have relevance to our work, and how we could build upon them in our future work.

## 2. MOTIVATION AND BACKGROUND
In this section, we discuss the existing problems faced by the grid sites[1] in the explicit context of ATLAS experiment that are core motivation for the work presented in this paper.

## 2.1 Grid Site Constraints
Following are the major issues:

### 2.1.1 Site Configuration
Whenever new middle ware patches are released for the grid sites, a number of sites get mis-configured resulting in the site going offline. This results in no jobs being submitted to the site for a certain time until a system administrator corrects the problem. This causes under-utilization of resources and unnecessary administrative overhead for the site administrators.

### 2.1.2 Platform Availability
Presently ATLAS recommends Scientific Linux CERN 3/4 operating system, which is derived from Red Hat Enterprise Linux with CERN specific modification, for its grid sites. This is inconvenient as it reduces choice for site providers whom might have a different OS preference due to local needs and requirements.

---

[1] All batch farms and clusters participating in the Grid are referred to as *grid sites*.

### 2.1.3 ATLAS Analysis Software
An ATLAS experiment job requires analysis software that is around 7GB in size. To support the grid sites, which do not have their own local installation, CERN provides this software through read-only Andrew File System (AFS) over the Internet. Under heavy load, this leads to higher latencies, which results in frequent job failures and could be very costly in terms of job re-submission overhead for the long running jobs.

### 2.1.4 Software Update
The update cycle for the ATLAS experiment software is around 15 days and requires downloading of hundreds of tarballs to the site. This process is repetitive and burdensome for both the site providers and the ATLAS experiment software managers as it is prone to misconfigurations.

Executing ATLAS jobs in VM will certainly enable the site administrators to manage their physicals resources more efficiently as virtualization decouples the *execution environment* from the host machines by turning them into generic compute resources. Further more, pre-configured and pre-certified VM images with ATLAS software, available locally at the site, will provide an elegant solution to the current deployment strategy that relies on error prone software distribution and update life cycle.

## 2.2 Pilot Architecture
We first describe the PanDA System infrastructure. PanDA System first obtains resource leases by submitting "pilot jobs" to the WMS. Once the pilot job have gone through the Grid's WMS, and were assigned to a particular site, then the local resource manager (LRM) *aka* Batch System schedules the job on the available resources according to its *site policy*. At this point, PanDA pilot have acquired a resource slot, and it goes through 4 different stages during the lifetime of a particular job as following:

### 2.2.1 Environment Preparation
After downloading the input data set, pilot job wrapper prepares the environment and sets the necessary variables on the worker node so that job execution can proceed.

### 2.2.2 Stage-in
After setting up the environment, the pilot job wrapper downloads the input dataset as required by the job from the Grid Storage Element (SE).

### 2.2.3 Job Execution
Once all pre-requisites are met, PanDA pilot job wrapper triggers *runJob* script to commence job execution. During the lifetime of the job execution, the *runJob* script periodically sends the status updates to the pilot wrapper, which in turn updates the PanDA server to keep it aware with the progress of job to let it know that the job is still running.

### 2.2.4 Stage-out
In this phase, the pilot job wrapper updates the results to the grid storage element if the job completed successfully other wise it sends the error logs to the PanDA server.

## 3. IMPLEMENTATION

We have implemented a proof of concept *Application Programming Interface (API)* that enables a guest application, such as pilot jobs, to request virtual machine on a compute node. We called it VIRM (**VI**rtualization **R**esource **M**anagement) API. Our approach differs from prior techniques [6, 10, 14, 16] in that we propose a different abstraction of services that any VM deployment engine must provide in the context of Grid application.

In our study, we relied on vGrid [11] as deployment engine that uses the REST protocol for communication but other implementations could use XML, Web Services or other technologies. It deals with the tasks such as for setting up logical volume partitions, decompressing an operating system image in it, and starting/stopping virtual machines.

The *VIRM* API aims to provide the following abstract service end-points:

- **request diskspace:** to request a writeable workspace for the application to execute its job with a specified operating system
- **mount/unmount diskspace:** to have access to the workspace so that application could customize it and then unmount it aka *contextualization*
- **start/stop virtual machine:** to start and stop virtual machine with-in the workspace
- **remove diskspace:** to remove the workspace logical volume once the job has been executed
- **send heartbeat:** the guest application sends a periodic heartbeat message to the VIRM API to keep the virtual machine alive

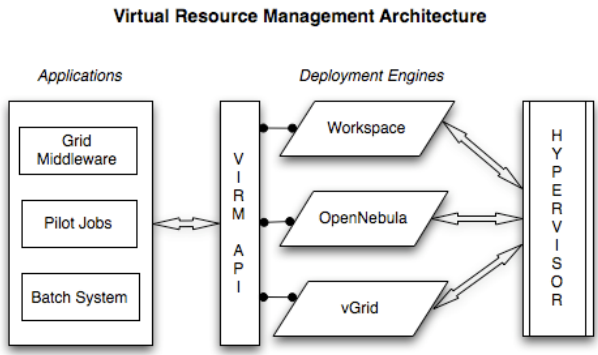

**Fig. 1. Applications access the virtualization capabilities available at the compute node through VIRM API that interfaces with a deployment engine such as VirtualWorkspace, OpenNEbula, vGrid etc and hides the lower level implementation details from the applications.**

To achieve above stated objectives, we modified the PanDA pilot framework and added interface end points for VIRM API, as shown in figure 1, which led to an over all architecture change [12].

At the deployment phase, once the pilot job has acquired a *resource slot* it verifies if *VIRM Service* is present over the loopback interface and upon valid response, it assumes that the job will be executed in the VM and follows the steps as shown in figure 2.

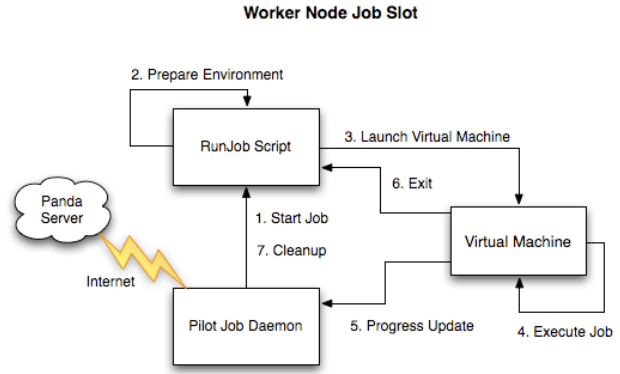

**Fig. 2. Once the pilot job has started, it launches the *runjob* script, which requests the virtual machine container from the *VIRM agent* and starts the job execution. Once started, the virtual machine updates the main pilot job of its status and upon job failing/termination; the *runjob* script requests the shutdown of the virtual machine.**

## 4. CASE STUDY

The biggest challenge for running HPC jobs in the VM on the Grid lies in how significant the virtualization overhead is, and how such a VM could be deployed on-demand using pilot jobs. To evaluate that, in this section we aim to answer following questions:

- How sensitive are HPC jobs to the amount of CPU processing resources allocated to the VM? Can an optimal performance be achieved while fitting in the maximum number of virtual machines on the server to achieve higher *resource utilization*?
- How significant the effect of scheduler parameters on the job performance i.e. weight and CPU cap for the Xen credit scheduler?
- Can performance be improved by outsourcing networking I/O to $Dom_0$[1] rather then $Dom_U$[2]?

Our test bed consist of two servers each with SMP dual-core Intel CPU's running at 3 GHz, 8 GB RAM and 1 Gbit/s NIC's running Xen 3.1 and Scientific Linux CERN 4 (SLC4). We used pre-configured locally cached SLC4 VM images with pre-installed ATLAS analysis software on logical volumes, which were attached to the VM as block-devices at the start-up time to create dynamic *virtual appliance* for ATLAS production jobs.

---

[1] $Dom_0$ is the admin domain on Xen, which is used for administrating the virtual machines.

[2] We use words *domain*, *virtual machine, VM* and $Dom_U$ interchangeably.

In this section, we present a comparative study of running ATLAS *reconstruction* jobs, which have high CPU, Memory and I/O requirements, on both physical and VM to benchmark their performance. Our results provide empirical evidence that optimization of Xen scheduling parameters have significant impact on the job performance. We used 10 different configurations under which we varied the number of allocated CPU's, memory, CPU weight and Cap (limit on percentage of CPU usage) relative to $Dom_0$ and to each parallel running $Dom_U$. $Conf\_1$ and $Conf\_2$ are our physical and virtual baselines against which all configuration results are compared.

Let $M_i$ donate the memory allocated to a $Dom_i$ running an ATLAS job with $P_i$ number of CPU allocated with a given Weight $W_i$. $K$ represents a constant time that it takes for setting up and shutting down the virtual machine of ~3mins which we measured in tests. Let $C_i$ as the CPU cap parameter for the Xen credit scheduler with $n_{VM}$ as load representing number of parallel $dom_i$.

$T_i$ donates as the estimate time it takes for the job to completed depending on the allocated values for the given parameters as illustrated in equation 1:

$$T_i = K + \frac{f(M_i) * f(W_i * P_i)}{C_i * n_{VM}} \quad (1)$$

## 4.1 Impact of CPU parameters on $dom_i$ performance

We first observe that a job's estimated time of completion depends on the amount of CPU allocated to it. This seems to be the most important factor in determining the job performance, see in Table 1, that when the CPU's allocation, $C_i$, was capped to 0.5, the time to complete the job almost increased by 83% which shows that the relationship between $T_i$ and $C_i$ is not one-to-one. It is important to be considered for more complex situations where diverse set of workloads may run in parallel $Dom_i$ each with different set of memory and CPU requirements, and poses a significant challenge for optimally allocating machine resources to competing $Dom_i$.

**Table 1. Time taken by the job to complete when different CPU number was allocated for each of 5 different configurations with 2GB of RAM for all $Dom_i$**

| Test # | $n_{VM}$ | $P_i$ | $T_i$ | Overhead (%) |
|---|---|---|---|---|
| Conf_1 | 0 | 4 | 7080 | 0 |
| Conf_2 | 1 | 4 | 7130 | 1 |
| Conf_5 | 2 | 2 | 7193 | 2 |
| Conf_9 | 3 | 1 | 7970 | 13 |
| Conf_10.1 | *3* | *0.5* | 12926 | 83 |

But if a single CPU is allocated to each $dom_i$ then the overhead (%) stays below 15%. This is very promising especially in case of $Conf\_9$ where 3 parallel $dom_i$ were run in parallel as compared to $Conf\_1$ physical baseline.

We also observe that *Fair-share* scheduling technique leads to a slightly better performance as compared to when CPU was pinned for each $dom_i$ as shown in figure 3(a) where larger set of physics events for workload got processed between 30s-50s.

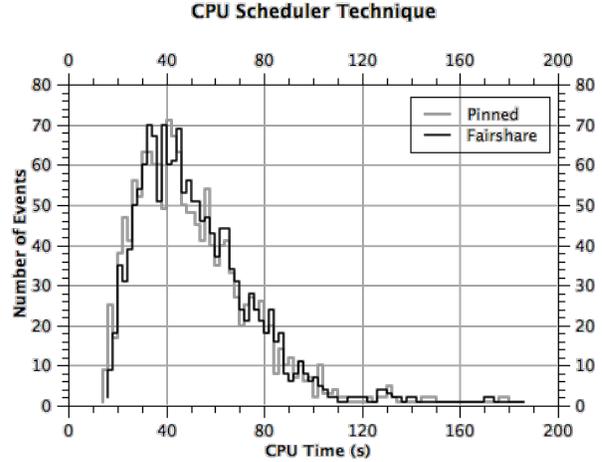

**Fig. 3a. Evaluating CPU scheduling parameters: Pinned vs Fairshare strategy for *$dom_i$***

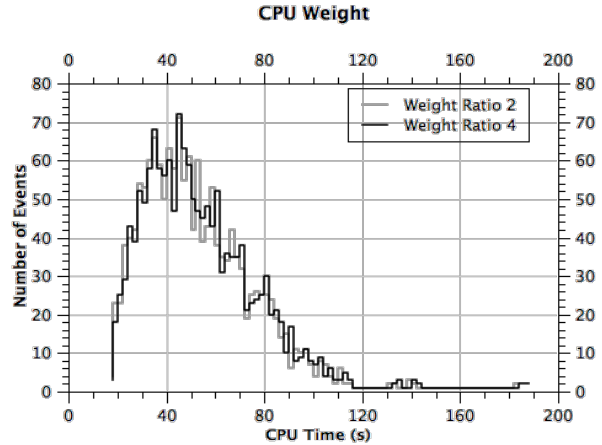

**Fig. 3b. Evaluating CPU scheduling parameters: Weight ratio strategy for *$dom_i$***

This may be due to the fact that pinning the CPU for each $dom_i$ constraints it even if other $dom_i$ may not be fully utilizing its CPU cycles. While Xen fair-share proportional scheduler gives a slice of 10ms to each $dom_i$ while shares are re-computed every 30ms leading to higher availability of CPU for any given $dom_i$.

Similarly, when we ran 2 parallel $dom_i$ with weight ratios, $W_i$, of 2 and 4 respectively as compared to $Dom_0$, we observed that varying $W_i$ among $dom_i$ only had a limited impact on the job performance where higher $W_i$ led to a slightly better performance. We intend to further investigate this in our future research by introducing different workload types in parallel $Dom_i$ building upon the work of Stiender et al [4].

## 4.2 Impact of Memory on $dom_i$ performance

Memory allocation $M_i$ to $dom_i$ is another important factor to be considered when an optimization of performance is required for particular type of workloads. The ATLAS workloads, in particular reconstruction of physics events, are highly memory intensive.

In our test suite, we progressively reduced $M_i$ available to $dom_i$ to observe the overall performance metric for the job. Figure 4 shows that reducing the $dom_i$ memory have a lesser performance impact when compared to number of parallel $dom_i$ each running CPU hungry tasks. $Dom_i$ with $M_i$ of 8GB, 7GB, 6GB, 3GB have only a variance of 100 sec but running 3 $dom_i$ each with 2GB for *Conf_9*, as explained in section 4.1, added additional 800 sec to the job completion.

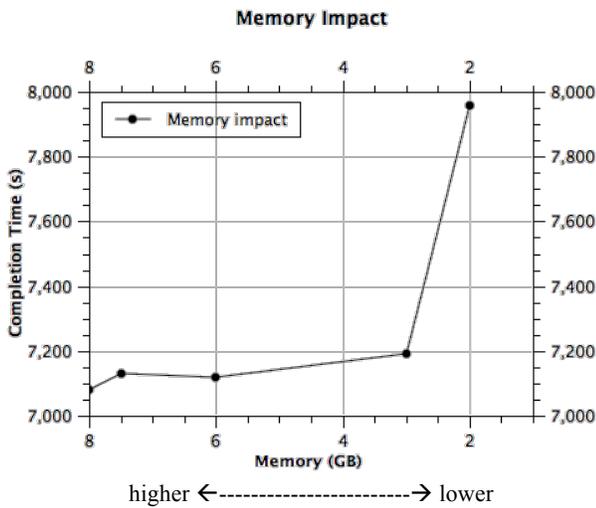

higher ←-------------------→ lower

**Fig. 4. Evaluating the impact of memory (RAM) on overall job completion time**

We also benchmarked the physical resident memory and virtual memory request patterns of our workloads in $Dom_i$. This is important for *reconstruction* jobs as they process large number of physics events, if the $Dom_0$ could not cope with those requests then it could terminate the job pre-maturely resulting in wastage to CPU resources which already have been consumed by the job.

Figure 5(a) and 5(b) provide an overview of the physical and virtual memory requests. Each job starts with an initialization phase, and as the events are processed the overall memory acquired-so-far increases. This actually reflects a memory leak in the application suite we used, that is not releasing memory acquired as the job progresses for the events that have been already processed. Fixing this bug is beyond the scope of our current study.

The only important difference to note is between *Conf_10.1* (with CPU cap set to 0.5) and *Conf_10.3* (with no CPU cap set) where both configurations were only given 2GB of RAM each. Setting the CPU cap results in slower release of memory to $Dom_i$ by the $Dom_0$ as it receives lesser CPU cycles which results in memory requests being queued up and processed only when the $Dom_i$ becomes active. Otherwise both configurations were able to acquire the memory resources they required eventually, and the workloads were executed successfully.

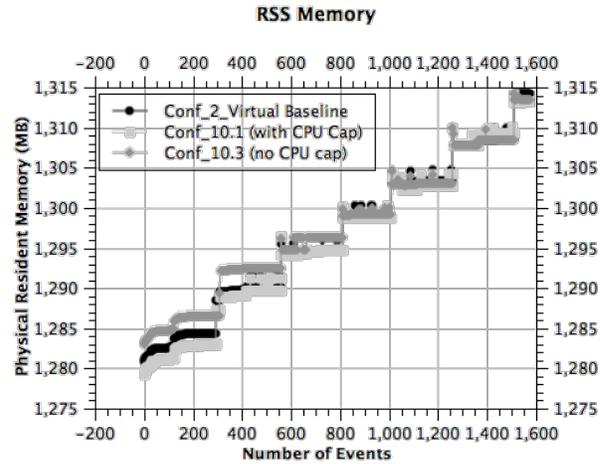

**Fig. 5a. Evaluating the impact of parallel running *domi* on physical memory requests**

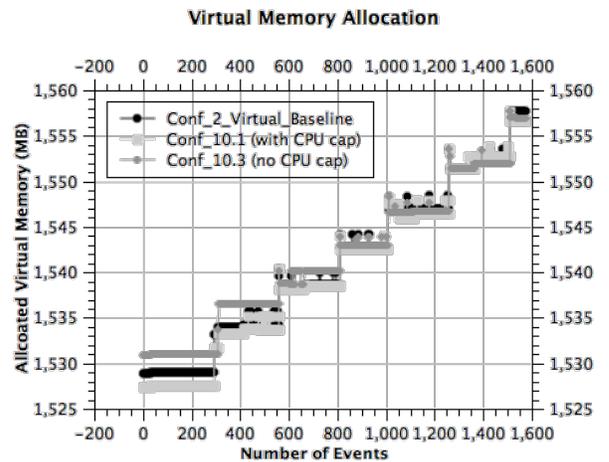

**Fig. 5b. Evaluating the impact of parallel running *domi* on virtual memory requests**

## 4.3 Network I/O performance of $dom_i$

Virtualization's largest overhead comes from the network I/O as Xen hypervisor uses CPU cycles to process network packets, thus leading to additional loss of performance. This is due to Xen's virtual network device driver architecture, which resides in $Dom_0$.

We observe, as shown in figure 6, that $Dom_0$ network throughput dropped significantly if constrained to 0.5 GB, but $Dom_0$ I/O throughput improved to ~50 Mb/s for the memory range of 1GB - 2GB or more while the average throughput achieved without Xen was ~65 Mb/s for peak and off-peak time. We ran our test using Linux *scp* (secure copy) utility on the LAN to avoid external network latencies interfering with our results.

This is a very significant factor especially for workload which requires large sets of input and output data files to be moved across the network as the job is executed especially for interactive jobs in $Dom_i$.

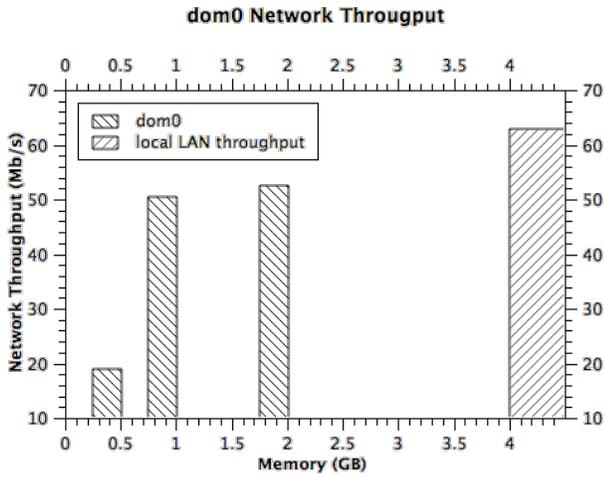

**Fig. 6. Evaluating the impact of memory on the through put of $dom_0$**

We then modified our test suite and ran again Linux *scp* utility to transfer a dataset of 3GB, on our LAN, for various configurations on $dom_i$. The results are presented in Table 2, and we found that $Dom_i$ avg. throughput was about ~6.5Mb/s when both source and the destination were $Dom_i$ deployed on two different servers. This scenario is important for Grid Storage Element (SE), as if virtualizing them could have drastic impact on the throughput of data transfers for various job requests.

**Table 2. Evaluation of network throughput for physical and parallel running $dom_i$ by transferring 3GB dataset**

| Parallel $dom_U$ # | Source-to-Destination | Throughput (Mb/s) |
|---|---|---|
| 0 | Physical-to-Physical | 62.8 |
| 0 | Physical-to-Virtual | 8.8 |
| 3 | Physical-to-Virtual | 8.3 |
| 3 | Virtual-to-Virtual | 6.4 |
| 3 | Virtual-to-Virtual | 6.6 |

The above results provide empirical evidence to prove that our approach to do all data upload and download in the $dom_0$ rather than $dom_i$ for the HPC workloads provides a significant boost to performance by removing network related virtualization over head from the equation. Our study also highlighted that reduced network throughput in $dom_i$ poses a considerable challenge for deploying Grid services on the virtual machines, and have to be further researched to fully understand its impact before grid services can be completely virtualized.

## 5. RELATED WORK

Our approach has been built upon the existing experience and know-how of the infrastructure such as the LHC[1] worldwide grid for the physics and scientific community, so that existing deployment frameworks such as the PanDA Pilot could be slightly modified to enable on-demand virtual machine based job execution on the Grid. This approach is different from projects like IN-VIGO [19] and DVC [5, 7], which focuses on creating ad-hoc "virtual computing grids" and presents them to the end user as grid sessions. In the modified PanDA Pilot, virtual machine based execution of the job stays transparent to the end user. Similarly, Maestro-VC [18] and Xenserver [17] focuses on a concept called *utility computing* where virtual resources could be leased and self-financed. Maestro-VC also explores how high performance parallel jobs could be deployed on on-demand virtual clusters.

On the other hand, since the past few years a number of projects such as Virtual Workspace [15], GridWay [13], OpenNEbula [14] have developed their own deployment engines, which could also be plugged into any batch system and cluster infrastructure to provide virtualization capability at the compute nodes in any grid. Our deployment engine, vGrid [11], is very similar to above-mentioned tools in capability and architecture, however it customizes the above approaches to serve the needs of the ATLAS community by providing not only resource management but also job management.

The challenge of using these deployment engines is that each of them have different interfaces and back-end software dependencies e.g. Virtual Workspace Service requires Globus Toolkit while OpenNEbula builds upon its own services. For any such services to be integrated into the scientific grids, there is a growing need for standards so that applications (PanDA Pilot, Virtual Workspace Pilot, Torque Batch system) could plug into them and tap into the virtualization capabilities. We attempt to address this standards gap in the *VIRM* API by abstracting basic set of core services that any deployment engine must provide. If such a common API is agreed and certified by Grid middle ware providers such as gLite [28] and Globus [29], similar to the GLUE schema for the information systems, to allow Pilot frameworks to deploy virtual machines on demand on Grid infrastructure, then site providers will have the freedom and choice to pick their preferred deployment engine as long as it adheres to agreed standards.

This will also open a whole new arena of diverse use cases where Pilot jobs could be used as a mechanism to setup dynamic and on-demand virtual clusters on the grid as envisaged by CERNVM[2] and their proposed CO-Pilot project. In one of our tests, we also evaluated CERNVM virtual machine image that have a network based file system and provides ATLAS analysis software over the network. We observed that there was an initial delay of ~3mins in downloading the required ATLAS software libraries for the job. Once the environment was configured, and then we ran tests with the same performance as in SLC4 images. This is very promising and effective way to customize the software environment according to the job requirements in the virtual machine as it boots up.

---

[1] Large Hadron Collider (LHC) is at CERN, Switzerland

[2] CERNVM project: http://cern.ch/cernvm

Our work is distinct and complementary to the Virtual Workspace Pilot by Freeman et al [16] in that rather than providing a new pilot framework, we have modified an existing Pilot framework that combined both resource provisioning and job execution. The PanDA Pilot framework is similar in functionality to the Virtual Workspace Pilot but has much wider scope in terms of the support for multiple batch systems such as Condor [27], Torque [26], LSF [25] and is in production on OSG, LCG Grids and NorduGrids.

## 6. CONCLUSION

In this paper, we have described *VIRM* API, an interface to enable pilot jobs and batch systems to seamlessly integrate Xen virtualization on the scientific grid and we presented an alternative deployment model for pilots where virtualization's network overhead is avoided by staging-in and staging-out job data outside of virtual machines. We also examined the impact of CPU scheduling strategies, memory allocation and network on the ATLAS experiments high performance computing jobs, and argued that virtualization, despite its overheads, provides very neat solutions to problems faced by the large experiments to effectively troubleshoot the grid sites when newer software is rolled out and to efficiently manage the application software environments.

Beyond CPU, I/O and memory performance, we wish to further develop our optimization model and expand it for concurrent HPC workloads of varying resource requirements on the same machine, and to explore that how efficient *resource utilization* could be achieved through dynamic scheduling of virtual machines.

We also plan to look into the more complex problems that how VM, running HPC workloads, could be dynamically scheduled, outside the scope of Grid Schedulers and Batch systems, at the server level by monitoring allocated CPU and memory to VM and then changing Xen scheduling parameters dynamically. We also wish to expand our future research on overall queue throughput if the jobs are executed only on VM in the particular context of ATLAS and other LHC experiment, and further developing VIRM API for Virtual Workspace and OpenNEbula platforms.

## 7. ACKNOWLEDGMENTS

We are grateful to ATLAS Computing people in providing support to us during our research and answering our long emails to explain especially by Dietrich Liko, Simone Campana, Alessandro De Salvo, Harry Ranshall, Alessandro De Girolamo, Benjamin Gaidioz and others. We also thank to Angelos Molfetas for his detailed feedback on the paper.

This work makes use of results produced by the Enabling Grids for E-sciencE (EGGE) project, a project co-funded by the European Commission (under contract number INFSO-RI-031688) through the Sixth Framework Programme. Full information is available at http://www.eu-egee.org.